\documentclass[aps,prd,preprint,superscriptaddress,showpacs,floatfix,nobibnotes]{revtex4-1}

\usepackage{latexsym}
\usepackage{amsmath}
\usepackage{amssymb}
\usepackage{graphicx}
\usepackage{color}
\usepackage{bm}
\usepackage{epsfig}
\usepackage{subcaption}
\newcommand{\bea}{\begin{eqnarray}}
\newcommand{\eea}{\end{eqnarray}}

\newcommand{\E}{{\mathcal E}}

\begin{document}

\title{Gradient Flow in the Ginzburg-Landau Model of Superconductivity}

\author{P. Mikula}
\email[]{pnmikula@gmail.com} \affiliation{Department of Physics, Brandon University, Brandon, Manitoba, R7A 6A9 Canada}
\affiliation{Department of Physics, University of Winnipeg, Winnipeg, Manitoba, R3B 2E9 Canada}
\affiliation{Department of Physics and Astronomy, University of Manitoba, Winnipeg, Manitoba, R3T 2N2 Canada}
\affiliation{Winnipeg Institute for Theoretical Physics, Winnipeg, Manitoba}

\author{M.E. Carrington}
\email[]{carrington@brandonu.ca} 
\affiliation{Department of Physics, Brandon University, Brandon, Manitoba, R7A 6A9 Canada}
\affiliation{Winnipeg Institute for Theoretical Physics, Winnipeg, Manitoba}

\author{G.~Kunstatter}
\email[]{gkunstatter@uwinnipeg.ca} 
\affiliation{Department of Physics, University of Winnipeg, Winnipeg, Manitoba, R3B 2E9 Canada}
\affiliation{Winnipeg Institute for Theoretical Physics, Winnipeg, Manitoba}

\date{\today}

\begin{abstract}
  We present numerical studies of the dynamics of vortices in the Ginzburg-Landau model using equations derived from the gradient flow of the free energy.
  Our equations are equivalent to the time dependent Ginzburg-Landau equations. 
   We have modeled the dynamics of multiple $n$-vortex configurations starting far from equilibrium. We find generically that there are two timescales for equilibration: a short timescale related to the formation time for a single $n$-vortex, and a longer timescale that characterizes vortex-vortex interactions.
  \end{abstract}

\pacs{11.10.-z, 
}

\normalsize
\maketitle

\normalsize

\section{Introduction}

The study of non-linear field equations that have particle-like solutions is of widespread interest in physics. Applications include confinement and superconductivity. An important example of a theory with classical soliton solutions is the Abelian Higgs model, which couples an Abelian gauge field $A_\mu$ to a charged (complex) scalar field $\phi$. The action for the Abelian Higgs model (with metric signature $(-+++)$) is:
\bea
S[A_\mu,\phi] = \int d^4 x \sqrt{g}\left[-\frac{1}{4} F^{\mu\nu}F_{\mu\nu} - (D^\mu \phi)^\dagger (D_\mu\phi) - V(\phi^\dagger\phi)\right] 
\eea
where 
\bea
F_{\mu\nu}&:=& \partial_\mu A_\nu - \partial_\nu A_\mu\\
D_\mu &:=&\partial_\mu +iq A_\mu\,.
\eea

The Abelian Higgs model forms the basis for the Ginzburg-Landau (GL) model \cite{GL} of superconductivity. Solutions to the GL equations extremize the free energy of the system, which is taken to be (minus) the three dimensional Euclidean  Abelian Higgs action. As is usually done, we assume translational symmetry in the $z$ direction, so that one is in effect studying a two dimension system. We work in natural units where $c=\hbar=\mu_0=1$ and the free energy is given by:
\bea
\label{action}
\E[A_i,\phi] = \int d^3x\,\sqrt{g}\,\big[\frac{1}{4} F^{ij}F_{ij}+(D^i \phi)^\dagger(D_i \phi)+V(\phi^\dagger\phi)\big] \,,
\label{eq:FreeEnergy}
\eea
where $\{i,j=1,2,3\}$, $g_{ij}=(+++)$ and the potential takes the form 
\bea 
V=\frac{\lambda}{4}(\phi^\dagger \phi-\frac{v^2}{2})^2 \,.
\eea
In the rest of this paper when we compute the energy density, we will for simplicity refer to it as the energy ($\E$), although it is really an energy per unit length. 

In the GL model, the temperature dependence is contained entirely in the vacuum expectation value for the scalar field, denoted $v(T)\equiv v$, and is given by:
\bea v = v(0)\left(1-\frac{T}{T_c}\right)^{-\frac{1}{2}}\,.
\eea
The two standard length scales in a superconductor, the coherence length ($\xi$) and the London penetration depth ($\Lambda_L$) are written in terms of the vacuum expectation value as:
\bea
\xi = \frac{2}{v\sqrt{\lambda}}\,\,\text{and}\,\, \Lambda_L = \frac{1}{qv}\,.
\eea
These two length scales are temperature dependent, but the GL parameter $\kappa$ is given by the ratio of the two length scales and is therefore dimensionless and temperature independent. The parameter $\kappa$ is defined:
\bea
\label{kappa-def}
\kappa^2=\frac{\Lambda_L^2}{\xi^2} = \frac{\lambda}{4q^2}\,.
\eea

It was shown by Nielsen and Olesen that this model has classical vortex solutions \cite{Nielsen}. Exact solutions were obtained in \cite{deVega} for $\kappa=1/2$. In this case, the GL equations reduce to first-order Bogomolnyi equations \cite{Bo, JT, Stuart}. 
In the GL model, $\kappa^2_c=1/2$ is a critical value that divides two physically distinct regions in the phase space which correspond to type I superconductors ($\kappa<\kappa_c$) and type II superconductors ($\kappa>\kappa_c$).

 We comment that in physical systems the behaviour at the critical point is actually quite complicated, and very different states can appear. 
These systems can be studied using extensions of Ginzburg Landau theory \cite{Lukyanchuk,Vagov,Peeters-a,Peeters-b,Peeters-c,Peeters-d}. The near critical regime is therefore both interesting and very complicated. 

In bulk physical systems, the sub-critical and super-critical cases exhibit distinctly different behaviour, in agreement with the predictions of the GL theory.
In type I superconductors ($\kappa<\kappa_c$) there are no stable vortices. The magnetic flux is expelled via induced surface currents. 
In type II superconductors ($\kappa>\kappa_c$), there are stable solutions containing multiple vortices with flux number $n=1$. A lattice of flux vortices forms which is known as an Abrikosov lattice \cite{Abrikosov}.
The stability of these solutions was proven in the GL model by Gustafson \cite{Gustafson}. 

In mesoscopic superconductors, different structures can develop due to the competition between surface effects and vortex-vortex interactions, which can lead to strong confinement of magnetic flux. Stable vortex solutions with all flux concentrated in a single vortex with $n>1$ quanta are possible. We refer to these as $n$-vortices - they are sometimes called `giant' vortices in the literature. 
These solutions have been found theoretically \cite{Schweigert-1} and recently identified experimentally in systems of size approaching the scale of the Cooper pairs \cite{Kanda,Cren,Xu}.

In this paper we study numerically the evolution and interactions of GL vortices, and discuss the connection with superconductivity. 
We are particularly interested in configurations that are far from equilibrium, which can reveal interesting features of the non-linear dynamics that are not readily apparent in a perturbative analysis.

Since the GL theory is a phenomenological model that describes a superconductor in thermodynamic equilibrium, the issue of the approach to equilibrium is not straightforward. 
Various dynamical equations have been proposed for describing the evolution of vortex configurations which start far from equilibrium, including relativistic equations derived from the four dimensional Abelian Higgs model, non-relativistic Schrodinger equation, and
the time dependent Ginzburg-Landau (TDGL) equations. 
The equations we study are equivalent to the TDGL equations as originally proposed in \cite{Gorkov} and \cite{Schmid} and extensively used in the literature  \cite{Maki,Du,Gropp,Crabtree,Chapman,Sardella,Kato}. We first show that these equations are most naturally understood as the gradient flow of the free energy. The gradient flow approach describes dissipative (non-conservative) time evolution analoguous to that of a generalized a heat equation. It also allows for a careful discussion of the issue of gauge invariance.

Our flow equations (and the vortex solutions) are similar to the Gross-Pitaevski equations (which depend on the scalar field only) whose evolution was computed in \cite{Schroers}, as well as the Chern-Simons vortices studied in \cite{Manton}. 
The action is different in these cases, but both calculations produce similar behaviour (vortex repulsion and orbiting pairs of vortices, under different circumstances). Flow equations are useful generally in situations where the geometry is interesting and it is ultimately our goal to apply the method to a system in curved space time. In this paper, we begin with the simple example of the Abelian Higgs model in flat space. 

We study both vortex formation and vortex interactions. 
Numerically, we focus specifically on identifying the important physical timescales associated with these processes.
We find the existence generically of two distinct scales: the first, shorter scale, characterises the formation of a single $n$-vortex, and the second, longer scale, is associated with vortex-vortex interactions. Similar results were found in \cite{Milo} where vortex anti-vortex pairs were seen to form very quickly, and then interact on a much slower timescale. Our findings are related to vortex transport properties and could be interesting in the context of recent works on the use of superconducting digital devices in supercomputers \cite{natcom,natcom2,natcom3,natcom4,natcom5}.

The study of vortex interactions also gives information about the stability of different vortex solutions. 
In a region of the parameter space where the stable solution is a multiple 1-vortex Abrikosov lattice, one expects that an $n$-vortex initial configuration would be unstable to decay. On the other hand, if the $n$-vortex solution is the stable one, one would expect that multiple 1-vortices would attract each other. 
Our study could therefore provide information about the conditions under which giant vortex solutions are stable. 

In our numerical calculations of vortex interactions, we use a finite 2-dimensional box and initial configurations with vortex centers that are not close to the edges. 
The size of the box is chosen to be large enough that the final energy is not affected by its size. 
However, finite size effects play an important role because our boundary conditions on the charged scalar impose fixed winding number, and therefore do not allow for the evolution of surface currents. 
In the super-critical region ($\kappa>\kappa_c$), we find that $n$-vortices are unstable to decay into multiple vortices with flux number $n=1$, which is consistent with previous theoretical and experimental results for type II superconductors. 
In the sub-critical region ($\kappa<\kappa_c$) we find a stable $n$-vortex solution.
One can understand this as a consequence of our flux conserving boundary conditions. 
In a sense the superconductor expels the flux as much as possible by confining it to a single large $n$-vortex.

The paper is organized as follows. In the next Section we present the flow equations that we will be studying. Section \ref{polar-section} studies the simple case of single $n$-vortices located at the origin with exact axial symmetry. We find that the energy of an $n$-vortex relative to $n$ 1-vortices obeys a simple power law relation. The exponent is less than two, even for very large values of $\kappa$, which disagrees with the heuristic argument in the textbook \cite{SCText}.
In Section \ref{cartesian-section} we consider various configurations of axially symmetric vortices which are shifted so that their centers are not (necessarily) at the origin. We study the formation and interaction of vortices, with specific emphasis on the timescales related to these processes.
We end with summary and conclusions. A brief description of some numerical details is relegated to an appendix.

\section{Flow Equations}

In general, the action, or free energy, is a functional $S[\Phi^I]$ of a set of fields $\Phi^I$, where the  index $I$ labels all dynamical fields in the action and also denotes all indices, internal as well as space-time coordinates. We assume that the field space is endowed with a positive definite inner product:
\bea
\langle\delta\Phi|\delta\Phi\rangle &=& G_{IJ} \delta \Phi^I \delta\Phi^J \,.
\label{eq:InnerProduct}
\eea
The summation convention implies summation over all discrete indices as well as  invariant integration over space-time coordinates.  For example, in the case of the Abelian Higgs model, the inner product for the  vector potential is
\bea
\langle \delta A|\delta A\rangle  &=& \frac{1}{k^2_A}\int d^3x \sqrt{g} g^{ij}\delta A_i\delta A_j \,,
\eea
and for a complex scalar $\phi$ we have
\bea
\langle \delta \phi |\delta \phi \rangle &=& \frac{1}{k^2_{\phi}}\int d^3x \sqrt{g}\delta \phi^\dagger \delta \phi  \,.
\eea 
The constants $k_\phi$ and $k_A$ ensure the inner product is dimensionless.
Physically, $k_\phi$ is the diffusion constant and $k_A$ is the inverse conductivity (see equations (\ref{kato1}, \ref{kato2})). 
We will show below that the GL flow equations depend only on the ratio of these parameters. 

Given an action and inner product, the flow equations are defined as
\bea
\label{flow-def}
G_{IJ} \frac{d\Phi^I}{d\tau} = \frac{\delta S}{\delta\Phi^J}= -\frac{\delta \E}{\delta\Phi^J}
\eea
where $\tau$ is the flow parameter.
The geometrical form of the flow equations (\ref{flow-def}) ensures that they have the correct sign to take the form of dissipative heat equations, whatever the parametrization of the fields or choice of coordinates. 
In particular,
\bea
\label{heatequation}
\frac{d \E}{d\tau} = \frac{\delta \E}{\delta \Phi^J}\frac{d \Phi^J}{d\tau} =  - \frac{d \Phi^I}{d\tau}G_{IJ}\frac{d \Phi^J}{d\tau}\,.
\eea
Since the inner product  (\ref{eq:InnerProduct}) is positive definite, the free energy can only be decreased by the flow.

If a field configuration is chosen that exactly extremizes the free energy, the right side of (\ref{flow-def}) is identically zero.
To test the stability of a given solution, we can start with a field configuration that is close to a solution, but not necessarily very close, and solve the flow equation. If the  
trial configuration evolves away from the solution as $\tau$ increases, then the solution is unstable.
In situations where we cannot solve the equations analytically, we can find solutions numerically by solving the flow equations to determine field configurations that are stable as the parameter $\tau$ approaches infinity.

The flow equations (\ref{flow-def}) have the form:
\bea
\frac{\partial A_i}{\partial\tau} = \frac{k^2_A g_{ij}}{\sqrt{g}}\frac{\delta {\mathcal E}}{\delta A^j} &=& -\frac{k^2_A}{\sqrt{g}}\left(-\partial^j(\sqrt{g} F_{j i}) + i q  \left((D_i\phi)^\dagger\phi-\phi^\dagger D_i\phi\right)  \right)\\
\frac{\partial \phi}{\partial\tau} = \frac{k^2_{\phi}}{\sqrt{g}}\frac{\delta {\mathcal E}}{\delta \phi^\dagger} &=& k^2_{\phi}\left[\left(D_j D^j\phi\right)-\frac{dV}{d\phi^\dagger}\right]\\
\frac{\partial \phi^\dagger}{\partial\tau} =\frac{k^2_{\phi}}{\sqrt{g}}\frac{\delta {\mathcal E}}{\delta \phi} &=& k^2_{\phi}\left[ \left(D_j D^j\phi\right)^\dagger-\frac{dV}{d\phi}\right] \,.
\eea
We define dimensionless variables as follows. 
We consider all lengths in units of the coherence length $\xi$, that is, we write
\bea
&& X = \tilde X \xi \,,~~~\text{or} \,~~~ \tilde X = \frac{v\sqrt{\lambda}}{2} X\,,~~~X\in\{x,y,z,\rho\} \,.
\eea
The scalar field is written in terms of its vacuum expectation value: 
 
\bea
\label{phi-cartesian}
\phi=\tilde \phi v = \frac{v}{\sqrt{2}}(\tilde\phi_1+i \tilde\phi_2)\,,
\eea
and the gauge field is scaled so that 
\bea
\label{gf-scaling}
g^{ij} A_i A_j = \frac{v^2 \lambda }{4 q^2} \tilde g^{ij} \tilde A_i \tilde A_j \,.
\eea
In Cartesian coordinates this can be written as
\bea
A_i = \frac{v\sqrt{\lambda}}{2q} \tilde{A_i} = \xi H_{c2} \tilde{A_i}
\eea
where $H_{c2}$ denotes the upper critical field of the superconductor, using the notation of \cite{Kato}.

The flow equations are not invariant under time dependent gauge transformations. However, we can replace the $\tau$ derivatives with the appropriate gauge invariant derivatives
\bea
\frac{\partial \vec{\tilde A}}{\partial \tau} \rightarrow \frac{\partial \vec{\tilde A}}{\partial \tau} - q\vec\nabla Q ~~~\text{and}~~~
\frac{\partial \tilde\phi}{\partial \tau} \rightarrow \frac{\partial \tilde\phi}{\partial \tau} +i q\, Q\tilde\phi \,,
\eea
where $Q$ is a scalar that transforms as
\bea
Q \rightarrow Q - q\partial_\tau \chi
\eea
under the gauge transformation given in (\ref{gtranf}). 
$Q$ can be interpreted as the scalar potential $A_0$. 
In section \ref{polar-section} we consider an axially symmetric situation
that allows us to work with explicitly gauge invariant quantities. In section \ref{cartesian-section} we use a two dimensional Cartesian coordinate system and work in the gauge where $Q=0$. 
In both cases the flow equations are equivalent to the usual TDGL equations for superconductors as proposed in \cite{Gorkov} and \cite{Schmid}. 

We rewrite the flow parameter in terms of a dimensionless variable and a characteristic timescale which is defined as 
\bea
\label{t0-def}
t_0 = \frac{\Lambda_L^2(0)}{k_A^2} = \frac{\Lambda_L^2}{k_A^2} \frac{v^2}{v(0)^2}\,.
\eea
The rescaled flow parameter is denoted $t$ and defined
\bea
\tau = t\; t_0 \frac{v^2(0)}{v^2}.
\eea
When working in terms of the rescaled fields and units, it will also be useful to define a rescaled version of the free energy:
\bea
\E = \tilde\E v^2\,.
\eea
The fully rescaled flow equations have the form 
\bea
\label{phi1-flow}
&&\frac{\partial \tilde\phi_1}{\partial t} = \frac{k_\phi^2 \kappa^2}{k_A^2}\left[ \nabla^2\tilde\phi_1- \tilde\phi_1 (\tilde\phi_1^2 + \tilde\phi_2^2 -1) -\vec{\tilde A}^2 \tilde\phi_1 -2\vec{\tilde A}\cdot \vec\nabla\tilde\phi_2-\tilde\phi_2\,\vec\nabla\cdot \vec{\tilde A}\right] \\[2mm]
\label{phi2-flow}
&& \frac{\partial \tilde\phi_2}{\partial t} =\frac{k_\phi^2 \kappa^2}{k_A^2}\left[\nabla^2\tilde\phi_2- \tilde\phi_2 (\tilde\phi_1^2 + \tilde\phi_2^2 -1) -\vec{\tilde A}^2 \tilde\phi_2 +2\vec{\tilde A}\cdot \vec\nabla\tilde\phi_1+\tilde\phi_1\,\vec\nabla\cdot \vec{\tilde A}\right] \\[2mm]
\label{Avec-flow}
&& \frac{\partial\vec{ \tilde A}}{\partial t} = \kappa^2\left[\nabla^2 \vec{\tilde A}-\vec\nabla(\vec\nabla\cdot\vec{\tilde A})\right] - \left[\tilde\phi_1\,\vec\nabla\tilde\phi_2 - \tilde\phi_2\,\vec\nabla\tilde\phi_1 + \vec{\tilde A}(\tilde\phi_1^2+\tilde\phi_2^2)\right]\,.
\eea
The parameters  $k_\phi$ and $k_A$ are related to the normal state diffusion constant, $D$, and the conductivity, $\sigma$, determined from BCS theory as follows:
\bea
\label{kato1}
&& D = k_\phi^2   = \frac{\xi^2(0)}{12 t_{GL}}\,, \\
\label{kato2}
&& \sigma^{-1} = k_A^2 = \frac{\Lambda^2_L(0)}{t_{GL}} \,.
\eea
Equations (\ref{phi1-flow} - \ref{Avec-flow}) depend only on the ratio of $k_\phi$ and $k_A$, and from (\ref{kappa-def}) we obtain immediately $k_A^2/k_\phi^2 = 12\kappa^2$. 
Our equations (\ref{phi1-flow} - \ref{Avec-flow}) correspond to Eq. (8) of \cite{Kato}, without the random force term, and identifying $t_0$ in equation (\ref{t0-def}) with the Ginzburg-Landau timescale $t_{GL}$, in the notation of \cite{Kato}. 
Also, since we work with fixed temperature $T$, we have rescaled by the temperature dependent quantities, rather than their values at $T=0$.


\section{Axially symmetric vortices}
\label{polar-section}

In this section we work in cylindrical coordinates and use an axially symmetric ansatz that reduces the problem to a 1-dimensional calculation. 
We parametrize the complex scalar field using two real functions which correspond to the magnitude  and phase of the complex field.
We define the components of the scalar field as
\bea
\label{phi-polar}
\frac{\phi}{v}=\tilde\phi =\frac{f}{\sqrt{2}}\,e^{i \omega}\,.
\eea
We use the ansatz 
\bea
\label{ansatz}
A_\rho=0\,,~~~A_z=0\,,~~~\tilde A_\theta=\tilde A_\theta(\tilde\rho)\,,~~~f = f(\tilde{\rho})\,~~~ \omega = - n \theta\,,
\eea
and therefore (\ref{gf-scaling}) with 
\bea
 \tilde g^{\theta\theta} = \frac{1}{\tilde \rho^2} \, ~~~~\text{gives}~~~~\, \tilde A_\theta = q A_\theta.
\eea

We define a new field $B$ as
\bea
\label{B-def}
B=-(\tilde A_\theta + \partial_\theta \omega\,) \,=-(\tilde A_\theta - n\,).
\eea
In order for the field $\phi$ to be single valued we require $n$ to be an integer, which is known as the winding number. The free energy in terms of these variables is given by
\bea
\label{eng-dif}
\tilde\E = \pi \int d\tilde{\rho} \,\left[\frac{1}{\tilde{\rho}} (\kappa^2 B^{\prime 2}+f^2 B^2) + f^{\prime 2} \tilde{\rho} +\frac{\tilde{\rho}}{2}(f^4-2f^2+1)\right]\,,
\eea
where we have performed the integration over $d\theta$, and the flow equations have the form 
\bea
\label{B-flow}
&& \frac{\partial B }{\partial t} = \kappa^2\left(B^{\prime\prime}-\frac{B^\prime}{\tilde\rho}\right) - f^2 B\,, \\
\label{f-flow}
&& \frac{\partial f}{\partial t} = \frac{1}{12}\left[ f^{\prime\prime}+\frac{f^{\prime}}{\tilde\rho}-\frac{f B^2}{\tilde\rho^2}- f(f^2-1)\right]\,,\\
\label{omega-flow}
&&\frac{\partial \omega }{\partial t} = \frac{-1}{\tilde\rho}\partial_\theta \left(\frac{f B}{\tilde\rho}\right)\,.
\eea
where the prime denotes differentiation with respect to $\tilde\rho$.
Using (\ref{ansatz}) the right side of (\ref{omega-flow}) is zero, and therefore $\omega$ is constant along the flow.

The energy is invariant under a transformation which has the form in tilde variables
\bea
\label{gtranf}
\tilde \phi \to e^{i q \chi}\tilde\phi \,,~~\tilde A_\theta \to \tilde A_\theta -q\partial_\theta \chi\,.
\eea
Using the axially symmetric ansatz  (\ref{ansatz}) this transformation takes the form
\bea
\omega \to \omega + q \chi\,,~~\tilde A_\theta \to \tilde A_\theta-q\partial_\theta \chi
\eea
which shows that the field $B$ in (\ref{B-def}) is gauge invariant. 
In the rest of this section, we suppress the tildes. 

It is easy to show that (\ref{B-flow}) and (\ref{f-flow}) can be rewritten as first order equations of the form
\bea
\label{f1}
&& f^\prime-c_1 \frac{B f}{\rho} =0 \\
\label{B1}
&& \frac{1}{\rho}B^\prime -c_2(f^2-c_3)=0
\eea
where the parameters $c_i$ are constants.
Differentiating (\ref{f1}) and (\ref{B1}) and rearranging one finds that the resulting second order differential equations are equivalent to the original equations (\ref{B-flow}, \ref{f-flow}) if 
\bea
c_1=\pm 1\,,~~c_2=\pm 1\,,~~c_3= 1\,,~~\kappa^2=\frac{1}{2}\,.
\eea
We choose the upper (positive) solution in order to obtain finite energy, as explained below. 
We can obtain a decoupled second order equation for $B$ if we differentiate (\ref{B1}), and then use (\ref{f1}) to eliminate $f^\prime$ and (\ref{B1}) to remove $f^2$. The result of this procedure is the set of equations
\bea
&& B^{\prime\prime}+2\frac{B B^\prime}{\rho}-\frac{B^\prime}{\rho}-B = 0\,,\\
&& f^2=1-2\frac{B^\prime}{f}\,, 
\eea
which can be solved analytically \cite{deVega}.

In order to solve the general flow equations, we must choose the initial configurations from which to start the flow. 
We require these configurations to give finite energy, which restricts the behaviour of the fields in the limit $\rho\to\infty$. 
We also want the flow to preserve the form of the initial configurations at $\rho\ll 1$. We discuss below how to choose boundary conditions so that these conditions are satisfied. 

A pure superconducting state given by
\bea
\label{pure-sc}
f = 1\,, \qquad \; \vec A = 0 \qquad \text{and} \qquad n = 0
\eea
has zero free energy (from equation \eqref{eng-dif}). If the fields approach (\ref{pure-sc}) as $\rho\to \infty$ then the energy will be finite. We therefore use the boundary condition 
\bea
\label{form-asy}
\lim_{\rho\to\infty} f\to 1\,,~~ \lim_{\rho\to\infty}B\to 0\,.
\eea

In the small $\rho$ limit we assume an expansion of the form 
\bea
\label{B-exp}
B = b_0+b_2\rho^2+b_4 \rho^4+\cdots\\
\label{f-exp}
f = \rho^\gamma(a_0+a_2\rho^2+a_4 \rho^4+\cdots)\,.
\eea
Substituting (\ref{f-exp}) into the right side of (\ref{f-flow}) we obtain 
\bea
\label{fdot}
\dot f = \rho^{\gamma-2} a_0(\gamma^2-b_0^2) + \rho^\gamma \big(a_2((2+\gamma)^2-b_0^2)+a_0(\kappa^2-2b_0 b_2)+\big) + {\cal O}(\rho^{\gamma+2})\,.
\eea
If the flow preserves the boundary conditions, we need the forms of (\ref{f-exp}) and (\ref{fdot}) to be the same, which means that the coefficient of the term of order $\rho^{\gamma-2}$ must be zero, from which we have
\bea
\gamma^2 = b_0^2\,.
\eea
Substituting (\ref{B-exp}) into the right side of (\ref{B-flow}) gives
\bea
\label{Bdot}
\dot B = 8 \rho^2 b_4-\rho^{2\gamma}\big(a_0^2 b_0+a_0(2 a_2 b_0+a_0 b_2) \rho^2 +{\cal O}(\rho^4)\big)\,.
\eea
The consistency of (\ref{B-exp}) and (\ref{Bdot}) requires $\gamma$, and hence $b_0$ to be an integer greater than or equal to one. The natural choice is to take $b_0=n$ (see equation (\ref{B-def})). Equations (\ref{B-exp}) and (\ref{f-exp}) therefore give $B(0)=n$ and $f(0)=0$.

The results of the previous two paragraphs give the boundary conditions on the flow as
\bea
\label{boundary-conditions}
B(0)=n\,,~~f(0)=0\,,~~\lim_{\rho\to\infty}B(\rho) = 0\,,~~\lim_{\rho\to\infty}f(\rho) = 1\,.
\eea
We will solve the flow equations by choosing initial configurations for the fields that satisfy these conditions. We also require that the fields are constant in time at the spatial boundaries, which means that (\ref{boundary-conditions}) is enforced at every step $dt$ throughout the flow.
We start at $t=0$ with the configurations 
\bea
\label{initial-x}
B(\rho)=n e^{-\rho^2}+g_1(\rho) \\
f(\rho) = 1-e^{-\rho^{|n|}}+g_2(\rho) 
\eea
where the $g_i$ are arbitrary functions with compact support such that
\bea
g_1(0)=g_2(0)=0\,.
\eea
Without loss of generality we can consider $n\ge 0$, which gives vortices with flux in the positive $z$ direction. Negative vortices differ only by an overall sign. 

\begin{figure}[htb]
\begin{subfigure}[b]{0.49\textwidth}
\includegraphics[width=\textwidth]{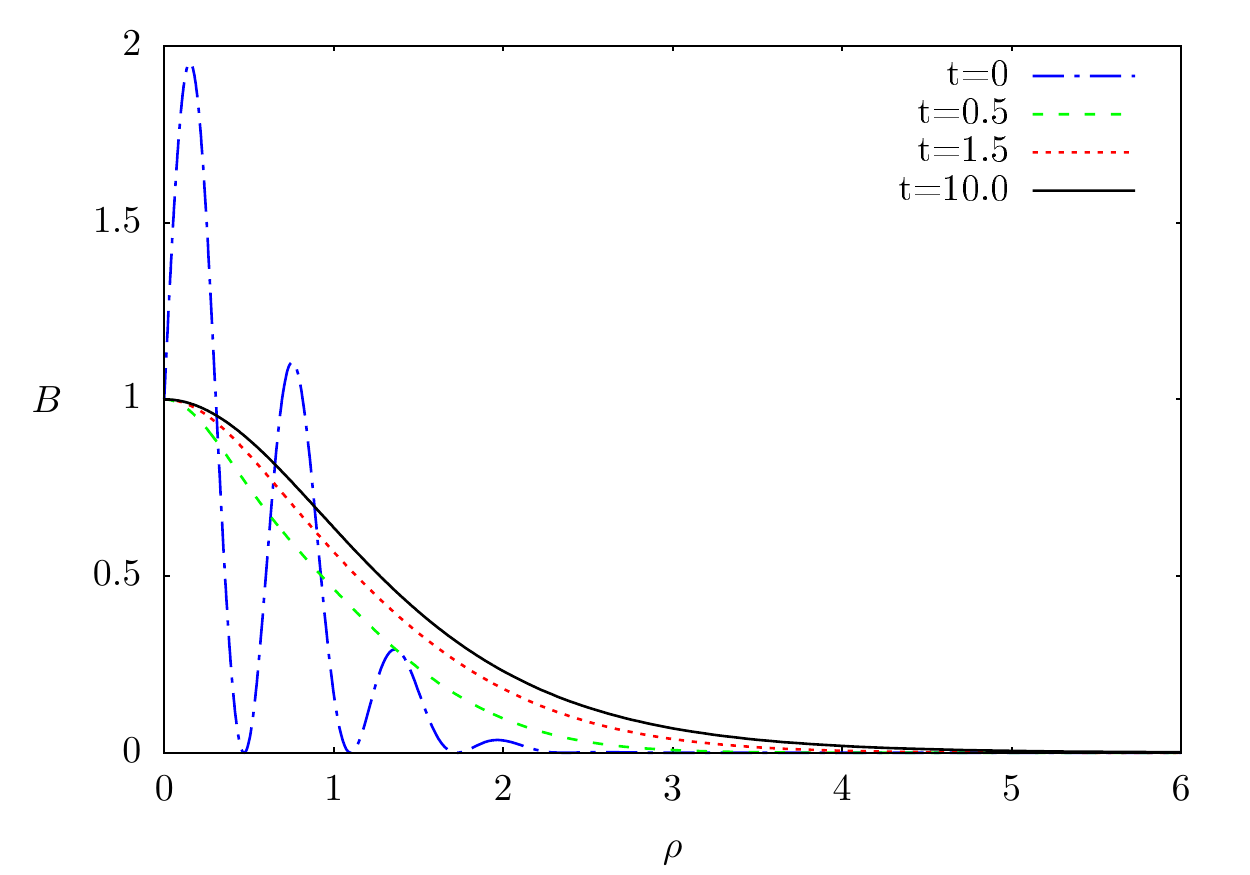}
\caption{$B(\rho)$ for $g_1=e^{-\rho^2}\sin(10\rho)$}
\end{subfigure}
\begin{subfigure}[b]{0.49\textwidth}
\includegraphics[width=\textwidth]{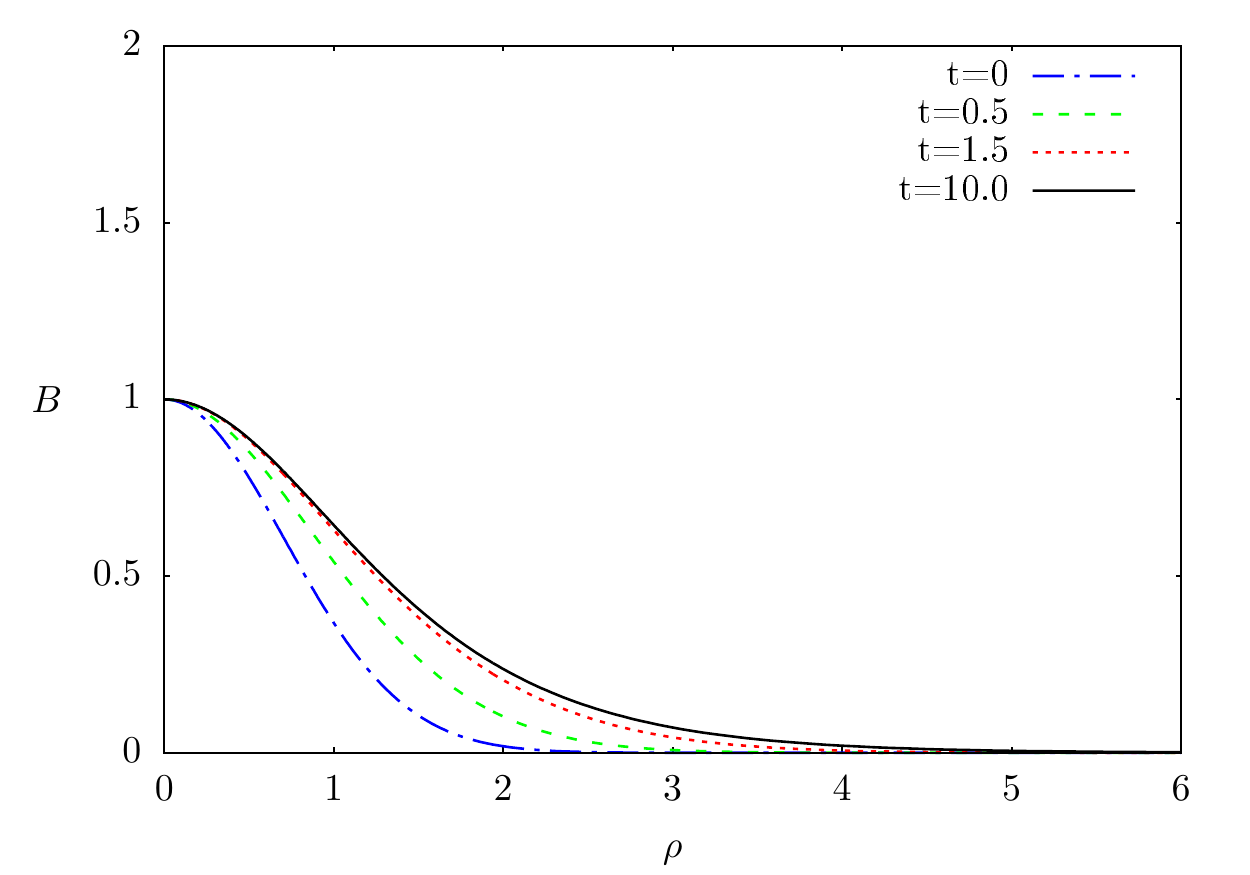}
\caption{$B(\rho)$ for $g_1=0$}
\end{subfigure}
\begin{subfigure}[b]{0.49\textwidth}
\includegraphics[width=\textwidth]{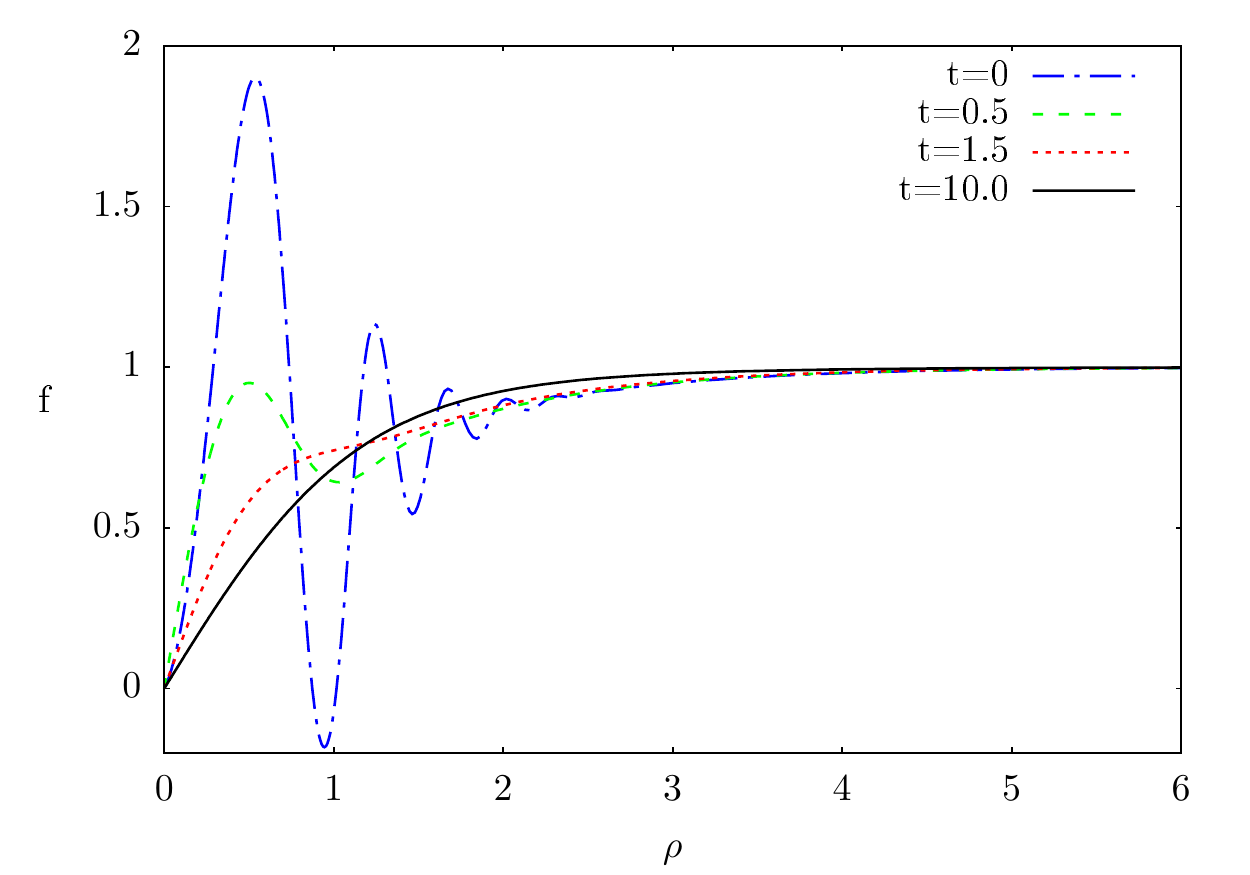}
\caption{$f(\rho)$ for $g_2=2e^{-\rho^2}\sin(5\rho^2)$ }
\end{subfigure}
\begin{subfigure}[b]{0.49\textwidth}
\includegraphics[width=\textwidth]{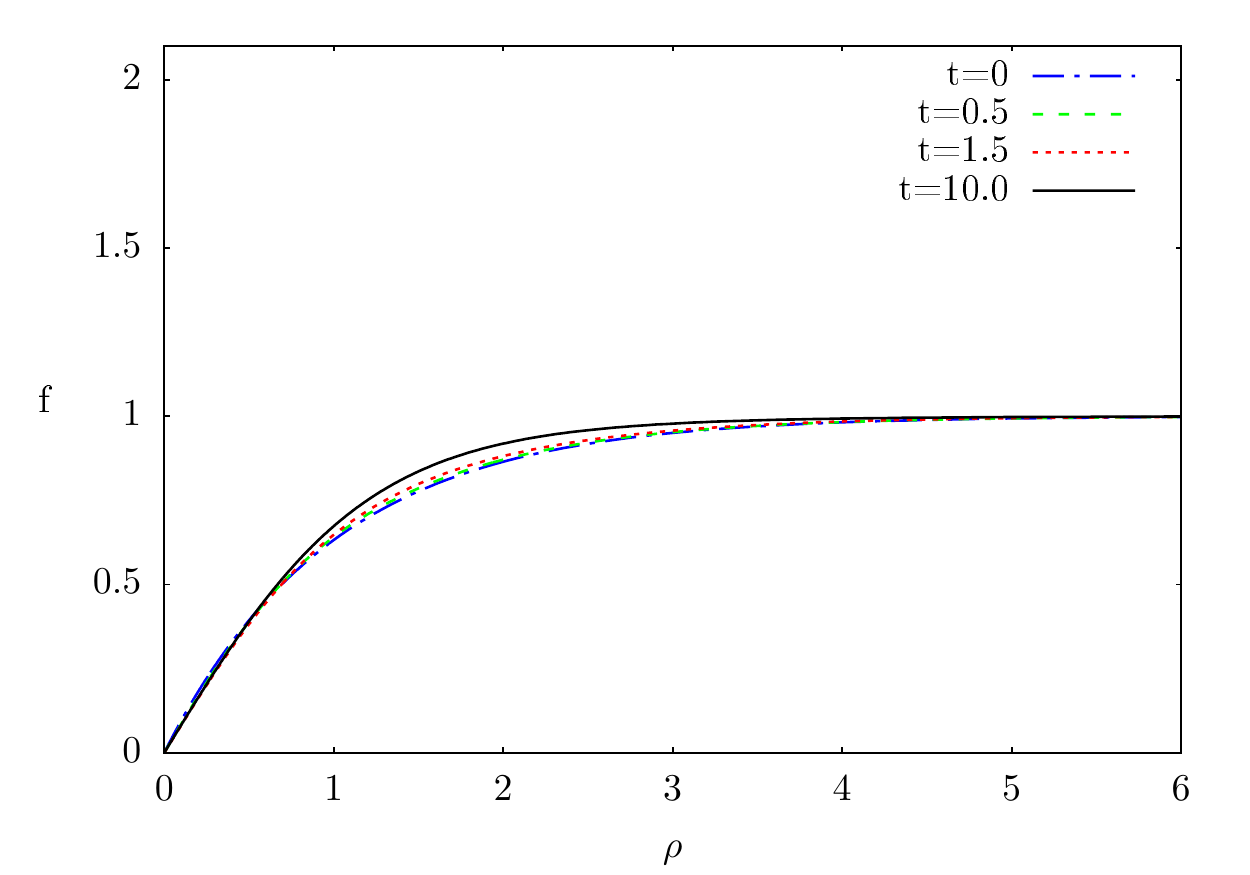}
\caption{$f(\rho)$ for $g_2=0$}
\end{subfigure}
\caption{At $t=10$ the flow with $\kappa=\kappa_c$ has converged to the solution for $n=1$. (Colour available online)}
\label{1dPlots}
\end{figure}

The constraint that the scalar field must be single valued means that the winding number must be an integer. Since the flow is continuous, the winding number does not change from its initial value during the flow (it is topologically conserved). The energy and flux do change under the flow but,
as shown in Fig. \ref{1dPlots}, both flow rapidly to their asymptotic values, with very little dependence on the choice of the initial configuration from which the flow begins. The same behaviour was found in Ref. \cite{Tang}.

To check our method we look at $\kappa=\kappa_c$, where the energy and flux of the vortices can be determined analytically \cite{deVega}:
\bea
\label{critical-energy}
\E_\infty = n \pi  \,,~~~~~~~~~~~~\Phi_\infty = 2\pi n\,.
\eea
The results are shown in Fig. \ref{1dPlots} for $n=1$. 
The asymptotic value of the flux is reached numerically in the limit of long flow times. The correct asymptotic value of the energy is also reached numerically.
We see from Fig. \ref{1dPlots} that even when the functions $g_i$ introduce large perturbations, there is little  effect on the time required to converge to the solution.

\begin{figure}
\includegraphics[width=0.8\textwidth]{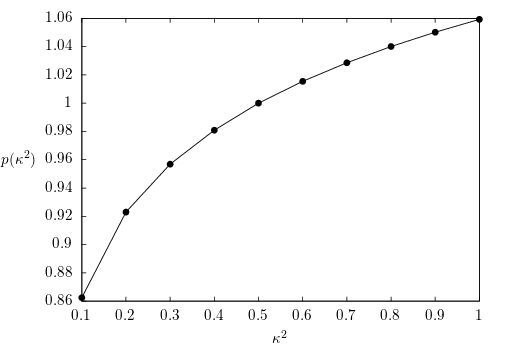}
\caption{The exponent $p(\kappa^2)$ of the power law defined in \eqref{Plaw}.\label{kappa-graph}}
\end{figure}

We also calculate the energy for  different values of $\kappa$ and $n$. The energy of the vortices for $\kappa$ on either side of the critical value is summarized in equation (\ref{kappa-summary}).
\bea
&& \kappa>\kappa_c~~~\to~~~ n \E_1 < \E_n \nonumber  \\
\label{kappa-summary}
&&\kappa=\kappa_c~~~\to~~~ n \E_1 = \E_n  \\
&&\kappa<\kappa_c~~~\to~~~ \E_n < n \E_1   \nonumber\,.
\eea
These results verify that for $\kappa>\kappa_c$ the system is unstable to decay into 1-vortices, and when $\kappa<\kappa_c$  the $n$-vortex is stable.
Fitting the data we find that the energy of an $n$-vortex relative to that of $n$ 1-vortices obeys a power law of the form,
\bea
\E_n = \E_1\,n^{p(\kappa^2)}\,.\label{Plaw}
\eea
The exponent $p(\kappa^2)$ is plotted in Fig. \ref{kappa-graph} for a range of $\kappa$'s in the neighborhood of the critical value. Note that each vortex energy is itself a function of $\kappa^2$ (the notation $\E_n$ means $\E_n(\kappa^2)$). 
At the critical value $p(\kappa^2_c)=1$, in agreement with \cite{deVega}. The function $p(\kappa^2)$ increases with $\kappa^2$ but is only weakly $\kappa^2$ dependent.
The approximate solution of \cite{Nielsen} predicts an $n^2$ dependence which agrees with the heuristic arguments of \cite{SCText}.
We find that the exponent increases slowly and does not appear to approach 2 in the asymptotic limit: for example, $p(1000)=1.58$.
Further discussion of the vortex energy and the static interaction energy between two vortices can be found in \cite{JacobsRebbi}.

We would like to study numerically the stability of $n$-vortices under arbitrary perturbations. 
This is impossible using axial symmetric vortices in polar coordinates, since any axially symmetric vortex in polar coordinates will be located at the origin. 
In order to study the stability of $n$-vortices, we consider the flow equations in a Cartesian coordinate system. This is the subject of the next section. 

\section{Two dimensional vortices}
\label{cartesian-section}

We would like to study the interaction of vortices at different locations. 
In this case there is no overall axial symmetry, and therefore it is no longer advantagous to work in polar coordinates. 
The reason is as follows: (1) in polar coordinates, the preferred center of an axially symmetric system is the origin, but we want mutiple distinct vortex centers; 
(2) an $n$-vortex is indistinguishable from $n$ 1-vortices, if all are located at the origin. To avoid this problem we work in 2-dimensional Cartesian coordinates. 
The flow equations are given in equations (\ref{phi1-flow}-\ref{Avec-flow}), and as in section \ref{polar-section} we will suppress the tildes. 

We consider axially symmetric vortices which can be shifted so that their centers are not at the origin. Since we work on a square lattice, the boundaries themselves break the axial symmetry, but we choose initial conditions so that the vortices are located far enough from the boundaries that the flow is unaffected by this asymmetry. A similar method was used in \cite{Schweigert-1,Schweigert-2}. 
The phase of the scalar field will not be a constant along the flow, as was the case using axial symmetry.
We want to start with a configuration for the scalar field that is single valued, so that we obtain an integer winding number which will then be conserved (topologically) by the flow. 
We therefore define the following scalar field configuration corresponding to a single vortex with winding number $n=1$ centered at the point $(x_i,y_i)$:
\bea
\label{phi-init-1}
\phi(x,y;x_i,y_i)= f(x,y;x_i,y_i)\,e^{i\omega(x,y;x_i,y_i)}
\eea
where 
\bea
\omega(x,y;x_i,y_i) = -\arctan\left(\frac{y-y_i}{x-x_i}\right)\,,~~~ \lim_{\vec r\to\infty}f(x,y;x_i,y_i)\to 1\,.
\eea
Parametrizing the complex scalar in terms of a real $(\phi_1)$ and imaginary $(i\phi_2)$ field equation (\ref{phi-init-1}) becomes
\bea
\label{initial-phi1}
\phi_1(x,y;x_i,y_i) &=& f(x,y;x_i,y_i)\,\cos\left(-\arctan\left(\frac{y-y_i}{x-x_i}\right)\right) \\
\label{initial-phi2}
\phi_2(x,y;x_i,y_i) &=& f(x,y;x_i,y_i)\,\sin\left(-\arctan\left(\frac{y-y_i}{x-x_i}\right)\right)\,,
\end{eqnarray}
and thus we have 
\bea
\lim_{\vec r\to\infty} \arctan \frac{\phi_1}{\phi_2} = -\arctan \frac{y}{x} ~~~\text{or}~~~ \lim_{\vec r\to\infty}\omega = -\theta\,,
\eea
which shows that the winding number is one (see equation (\ref{ansatz})).

In the two sections below we will consider two different situations: the formation of vortices, and the interaction of vortices. 

\subsection{Vortex Formation}
\label{vortex-formation-section}

\begin{figure}
\begin{subfigure}[b]{0.49\textwidth}
\includegraphics[width=\textwidth]{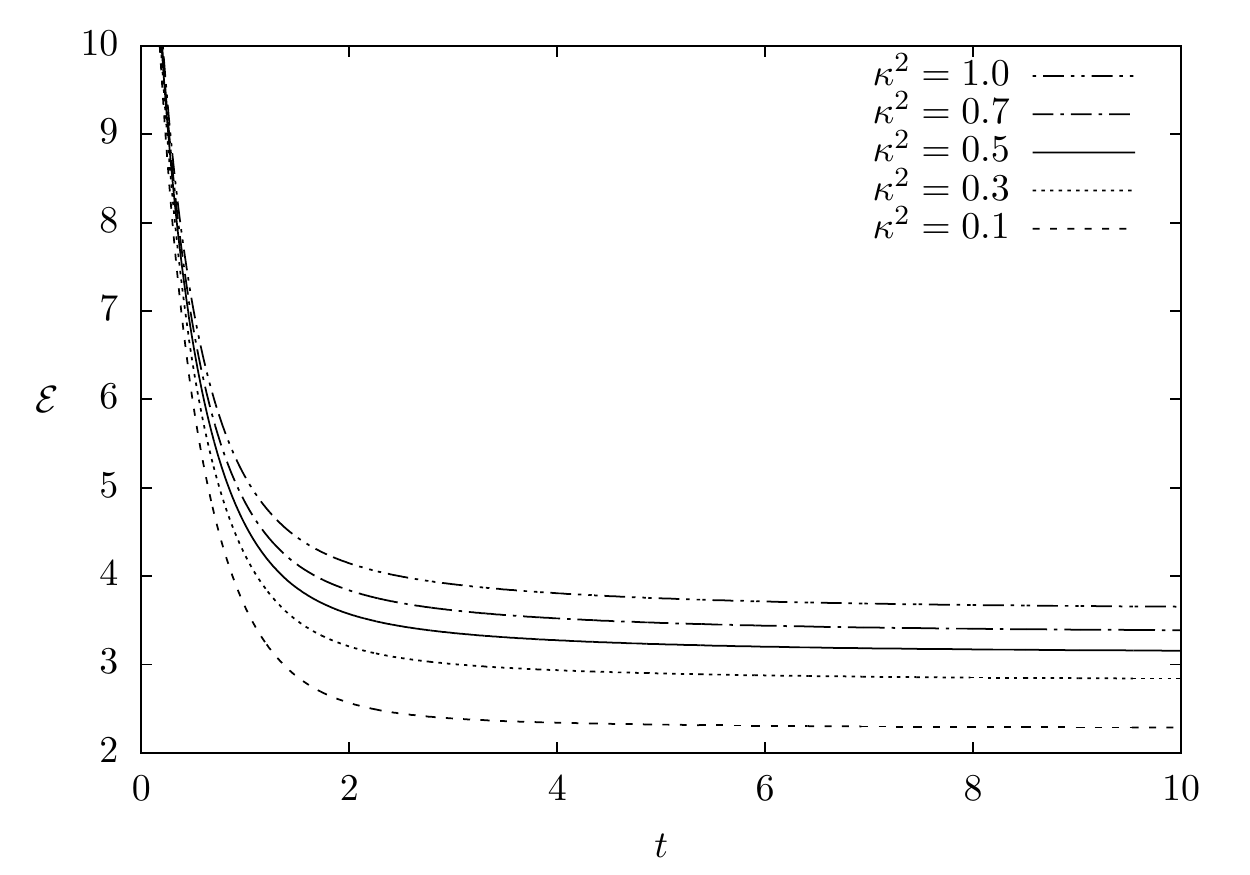}
\caption{\label{fig:EPlot}The energy of configurations where a 1-vortex forms at the origin. Larger values of $\kappa^2$ correspond to larger final energies.\\ \phantom{a}}
\end{subfigure}
\begin{subfigure}[b]{0.49\textwidth}
\includegraphics[width=\textwidth]{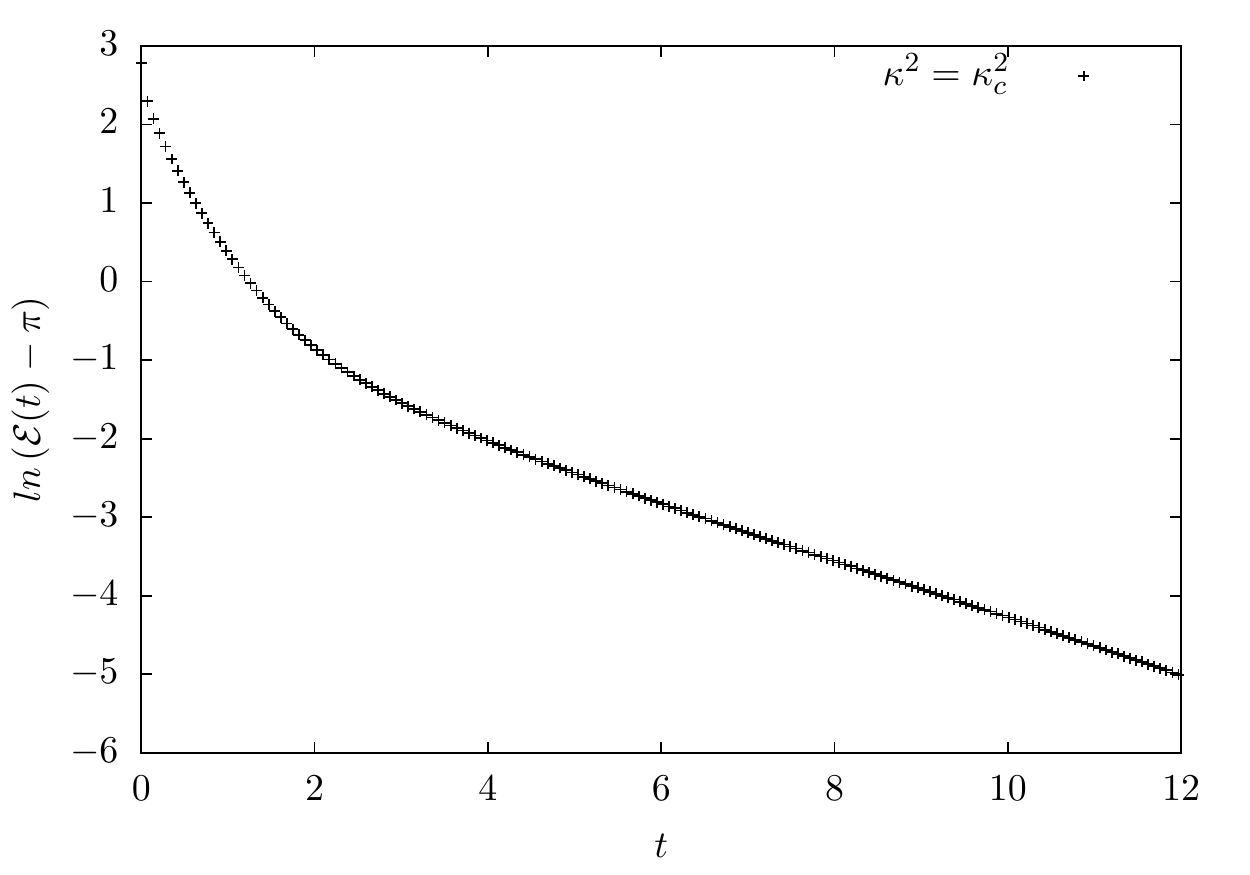}
\caption{\label{fig:logplot}$\ln[\E(t)-\pi]$ with $\kappa=\kappa_c$. The nonlinear behaviour is clear for early $t$, but the time dependence becomes linear as the flow progresses.}
\end{subfigure}
\caption{Energy of a configuration where a single vortex forms at the origin.}
\end{figure}

We initialize the scalar field using (\ref{phi-init-1}) centered at the origin with $f=1$ and $\omega=-\theta$. We set the vector potential (and magnetic field) to zero ($\vec A=0$) and evolve the configuration for various values of $\kappa$.
We calculate the evolution of the fields and the corresponding energy and flux. 
Note that the energy density in the $(x,y)$ plane of the initial configuration is constant, which leads to infinite energy if we consider the entire $(x,y)$ plane (see equations (\ref{eng-dif}) and (\ref{form-asy})). 
The initial energy is regulated by the size of the finite lattice in our numerical calculation. 
We have checked that the final energy is not affected by the finite size of the box, as long as the vortex centers are not close to the edges. We have also checked that when $\kappa=\kappa_c$ the energy approaches the correct finite value for a 1-vortex $\lim_{t\to \infty} {\E} = \pi$, as given in equation (\ref{critical-energy}).
In Fig. \ref{fig:EPlot} we show the energy as a function of the flow parameter for several values of $\kappa^2$. 
 The figure shows that the energy of a vortex increases with $\kappa$, but the evolution in time is similar for all values of $\kappa$. In Fig. \ref{fig:logplot} we see that $\ln(\E(t)-\pi)$ is nonlinear initially but becomes linear as the flow progresses. This suggests the existence of two distinct time scales in the vortex formation, a short time scale when the configuration is initially far from equilibrium and a longer time scale as the vortex settles to equilibrium.

\begin{figure}
\begin{subfigure}[b]{0.49\textwidth}
\includegraphics[width=\textwidth]{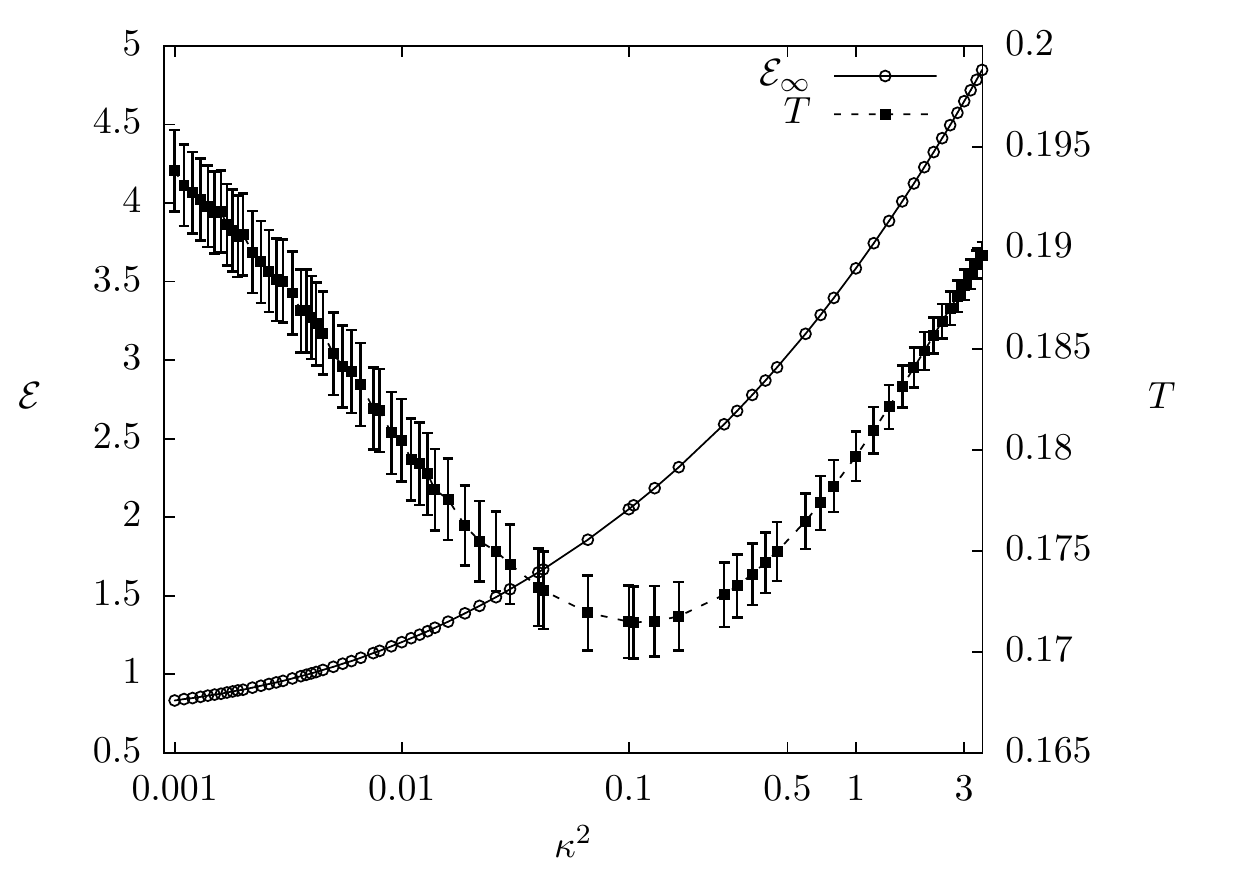}
\caption{\label{fig:ParaPlotsE}$T$ and $\E_\infty$  for $0<t<5$}
\end{subfigure}
\begin{subfigure}[b]{0.49\textwidth}
\includegraphics[width=\textwidth]{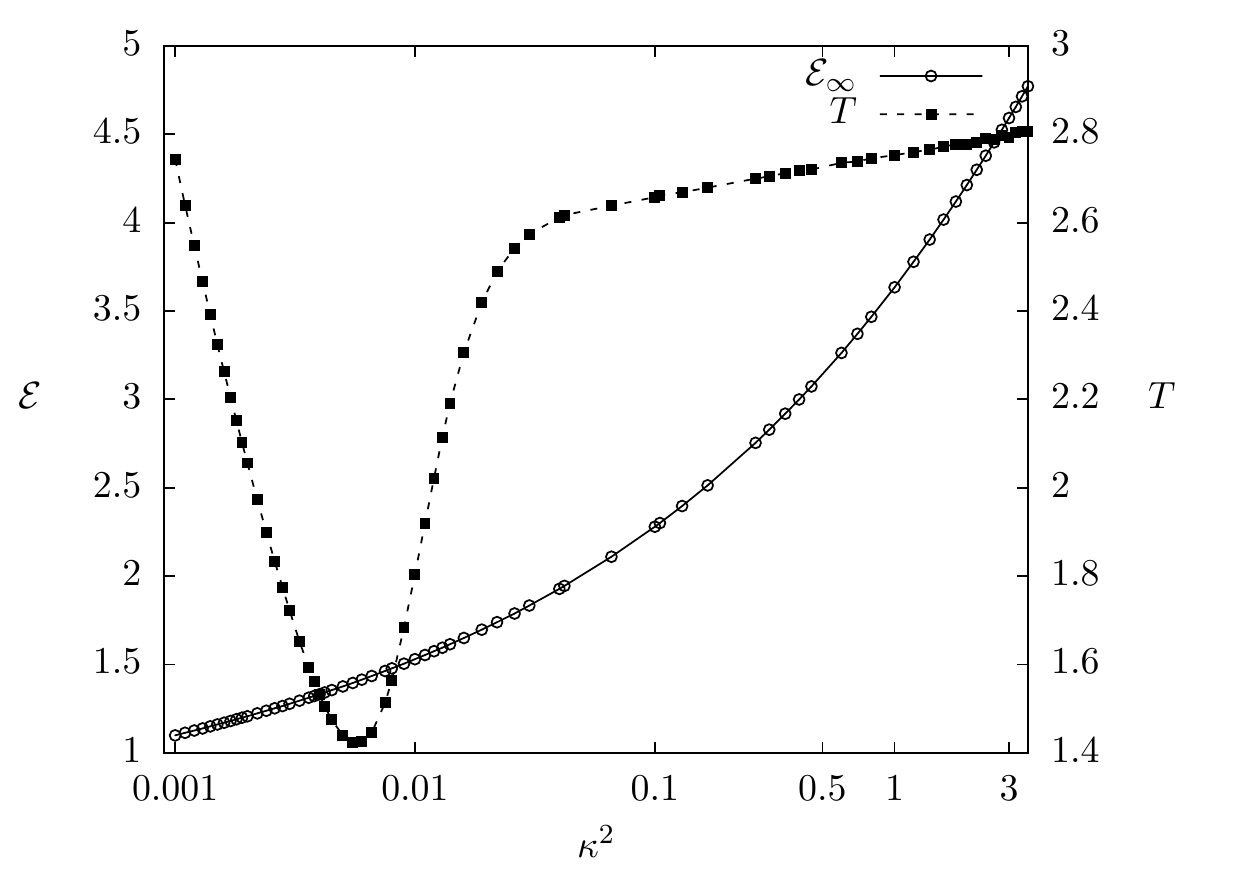}
\caption{\label{fig:ParaPlotsL}$T$ and $\E_\infty$ for $5<t<15$}
\end{subfigure}
\caption{Parameters obtained by fitting the data to equations \eqref{FitEqnEarly} in (a) and \eqref{FitEqnLate} in (b) for different values of $\kappa$.}
\end{figure}

In order to quantify the effect of $\kappa$ on the timescales for vortex formation, we fit the energy for $0<t<5$ to a function of the form
\bea
\E(t) = \E_\infty + (\Delta\E)e^{-\sqrt{t/T}} \label{FitEqnEarly}\,,
\eea
where we allow $\E_\infty$, $\Delta\E$, and $T$ to vary. For $5<t<15$ we fit to a function of the form 
\bea
\E(t) = \E_\infty + (\Delta\E)e^{-t/T} \label{FitEqnLate}\,.
\eea
The results are shown in Figs. \ref{fig:ParaPlotsE} and \ref{fig:ParaPlotsL}. The parameter $\E_\infty$ is a non-linear increasing function of $\kappa$ and the values obtained from the two fits agree within $3\%$. For early times, $T$ varies very little as a function of $\kappa$, but has a minimum near $\kappa^2=0.1$ as shown in Fig. \ref{fig:ParaPlotsE}. In Fig. \ref{fig:ParaPlotsL} we see that for later times $T$ has a minimum near $\kappa^2=0.0055$, and increases as $\kappa\rightarrow 0 $ or $\kappa\rightarrow\infty$. The graphs show the asymptotic standard error from the fits for the $T$ data points. 

We also considered the net magnetic flux $\Phi(t)$, starting from an intial configuration with zero flux $\Phi(0)=0$.  We expect that $\lim_{t\to \infty} {\Phi} = 2\pi$ (see equation (\ref{critical-energy})) since the magnetic flux of a vortex is independent of $\kappa$. In fact, we found that $\Phi(t)$ is independent of $\kappa$ to numerically accuracy at all points along the flow.

\subsection{Vortex Interactions}
To study the interaction of vortices we want to start with field configurations that are closer to vortex solutions, so that they can properly be considered vortices. 
We start once again with a discussion of a 1-vortex, but this time not centered at the origin. 
We choose initial conditions so that the magnitude of the scalar potential satisfies the condition $\lim_{\vec r\to\infty}f=1$, and the two components of the vector potential satisfy the flux quantization condition ($\Phi=2\pi $).
We define $r_i = \sqrt{(x-x_i)^2+(y-y_i)^2}$ and choose
\bea
\label{initial-f}
f(x,y;x_i,y_i)=\left(1-e^{-4\sqrt{\kappa^2}\,r_i^2}\right)\,
\eea
 and 
\begin{eqnarray}
\label{initial-A}
A_x(x,y;x_i,y_i) &=&  -(y-y_i)\left(\frac{1-e^{-r_i^2}}{r_i^2}\right)\\
A_y(x,y;x_i,y_i) &=&  (x-x_i)\left(\frac{1-e^{-r_i^2}}{r_i^2}\right) \nonumber
\eea
which gives a flux 
\bea
\Phi = \int_{-\infty}^\infty dx \int_{-\infty}^\infty dy \;(\partial_x A_y - \partial_y A_x) = \oint \vec A \cdot d\vec l = 2\pi\,,
\eea
where the line integral is taken around a square at infinity.
Numerically we use a finite range for the variables $x$ and $y$ but, because of the exponential factors in (\ref{initial-A}), $\Phi/(2\pi)$ is still approximately one as long as the vortex center is not close to the edges of the region of integration.

Now we want to construct the fields that correspond to a superposition of vortices  centered at different locations with winding numbers not necessarily equal to one. 
We use $(x^-_i,y^-_i)$ and $(x^+_i,y^+_i)$ to denote the initial coordinates of the centres of the negative and positive vortices respectively. 
In order to construct a superposition of 1-vortex configurations, the new vector field $\vec A$ is given by the sum of the component vector fields, while the new scalar field $\phi$ is given by the product of the component scalar fields. In this way, different gauge transformations acting on each of the vortices can be combined into a single gauge transformation of the composite fields. 
If we choose all $(x_i,y_i)$ distinct, the initial configuration corresponds to $n$ 1-vortices at different locations. 
If we choose coordinates so that two vortices are centered at the same location, we have effectively a 2-vortex at this position. 
We show below that the total winding number is equal to the sum of the component winding numbers.
Our initial configurations are
\bea
&& A_x = \sum_i^{n^+}A_x(x,y;x^+_i,y^+_i) - \sum_i^{n^-}A_x(x,y;x^-_i,y^-_i)\\
&& A_y = \sum_i^{n^+}A_y(x,y;x^+_i,y^+_i) - \sum_i^{n^-}A_y(x,y;x^-_i,y^-_i)\\
\label{initial-phi1-full}
&& \phi_1(x,y) = f\;\cos\left(\omega\right) \\
\label{initial-phi2-full}
&& \phi_2(x,y) = f\;\sin\left(\omega\right) 
\eea
where 
\bea
\label{omega-cond}
&& f = \prod_i^{n^+}f(x,y;x^+_i,y^+_i)\; \prod_i^{n^-}f(x,y;x^-_i,y^-_i) \\
&& \omega = \sum\limits_{i=1}^{n_-} \arctan\left(\frac{y-y^-_i}{x-x^-_i}\right) - \sum\limits_{i=1}^{n_+} \arctan\left(\frac{y-y^+_i}{x-x^+_i}\right)
\eea
and $\omega$ satisfies
\bea
\lim_{\vec r\rightarrow \infty} \omega =-(n_+-n_-) \theta\,.
\eea

We consider initial configurations with an $n$-vortex and a 1-vortex separated by a distance large enough that they do not significantly overlap. 
We find numerically that if $\kappa > \kappa_c$, the $n$-vortex decays into $n$ 1-vortices. This is shown in Fig. \ref{fig:Unstable}. 
For $\kappa < \kappa_c$ an $n$-vortex is stable, which means that the two vortices in the initial configuration will attract each other. 
In both the stable and unstable cases, the graph for the magnetic field has the same structure as the corresponding plots of the scalar field. 
At the critical value of the coupling $\kappa=\kappa_c$ the vortices do not interact, and the locations of the vortices do not deviate from those specified in the initial configuration. In this case we can find numerically stable solutions for any number of vortices, at any locations. \\

We would like to obtain some quantitative information about the timescales of vortex interactions.
We consider three different cases.

\begin{figure}
\begin{subfigure}[b]{0.49\textwidth}
\includegraphics[width=\textwidth]{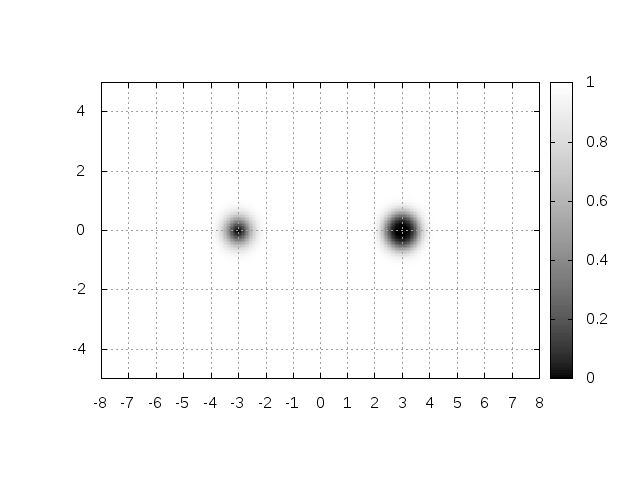}
\caption{The initial data for $|\phi |^2$ at $t=0$}
\end{subfigure}
\begin{subfigure}[b]{0.49\textwidth}
\includegraphics[width=\textwidth]{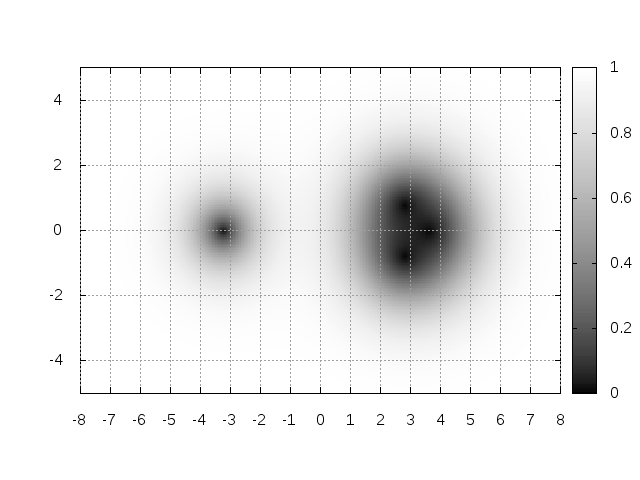}
\caption{$|\phi |^2$ after flowing to $t=100$}
\end{subfigure}

\begin{subfigure}[b]{0.49\textwidth}
\includegraphics[width=\textwidth]{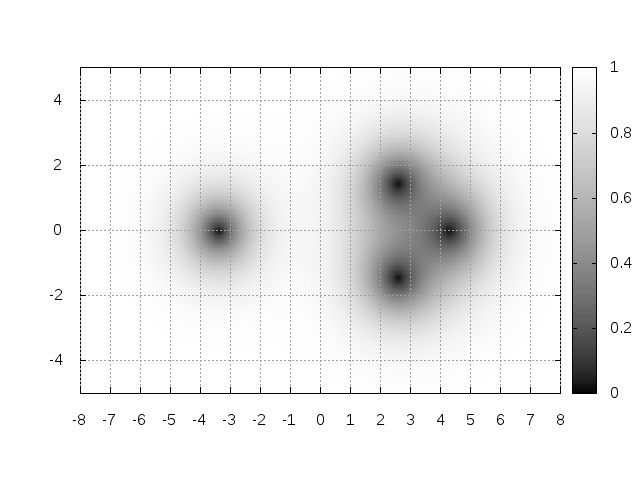}
\caption{$|\phi |^2$ after flowing to $t=200$}
\end{subfigure}
\begin{subfigure}[b]{0.49\textwidth}
\includegraphics[width=\textwidth]{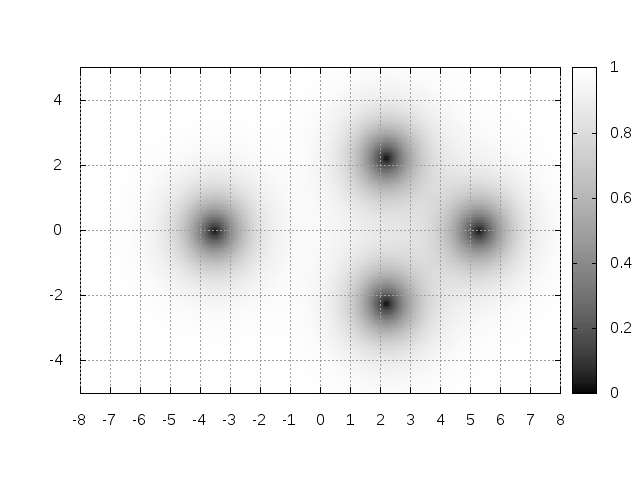}
\caption{$|\phi |^2$ after flowing to $t=320$}
\end{subfigure}
\rule{35em}{0.5pt}
\caption{\label{fig:Unstable} Decay of a 3-vortex initially located at $(3,0)$ with $\kappa=1$, in the presence of a 1-vortex at $(-3,0)$.}
\end{figure}

\subsubsection{Two 1-vortices}
\label{subsection-4-fig7}
In order to study the timescale of vortex interactions, we look at the interaction of two 1-vortices similar to the study 
in \cite{JacobsRebbi}. We find again the existence of two timescales with the best fit obtained using equation (\ref{FitEqnEarly}) for early times, and equation (\ref{FitEqnLate}) for later times. We discuss below the physical interpretation of these timescales. 

In section \ref{vortex-formation-section} we studied vortex formation and showed that in the early part of its evolution, a field configuration approaches equilibrium quite rapidly, with a characteristic timescale that is of order $0.2$ and  only weakly $\kappa$ dependent. This initial period of rapid change is followed by a longer period of slower evolution characterized by a timescale $\sim 2$. This separation of scales can be seen qualitatively from the early (steeper) and later (flatter) regions of the curve in Fig. \ref{fig:logplot}. 

A plot of $\ln[E(t)-E_\infty]$ versus $t$ for two approximate 1-vortices which start at (-1,0) and (0,1) is shown in Fig \ref{2scales-fig}. The curve shows the same separation into two steep/flat regions which correspond to different characteristic timescales. 
We will show below that the results from fitting to (\ref{FitEqnEarly}, \ref{FitEqnLate}) produce a typical short timescale that is the same order of magnitude as in section \ref{vortex-formation-section}, but the long timescale is orders of magnitude greater. 
We interpret these results by concluding that what we are seeing is a brief initial period during which the vortices form, followed by a longer phase that is dominated by the interaction of the formed vortices. 
The short and long timescales are therefore characteristic of vortex formation and vortex interactions, respectively. 
The key point is our assumption that the two timescales which are clearly seen in Fig \ref{2scales-fig} can be associated with the two distinct physical processes of formation and interaction. 
This assumption seems justified for four reasons: 
\begin{enumerate}
\item The simulations in section \ref{polar-section} showed that the timescale for vortex formation is largely insensitive to the initial configuration (see Fig. \ref{1dPlots}).
\item The short timescale has the same $\sqrt{t}$ dependence that was found in section \ref{vortex-formation-section} where vertex formation was studied.
\item The numerical values obtained for the short timescale are close to the numerical values for the short timescale for vortex formation from Fig. \ref{fig:ParaPlotsE}.
\item The interaction timescale depends strongly on $\kappa$ but the short timescale has a much weaker $\kappa$ dependence,  in agreement with Figs. \ref{fig:ParaPlotsE} and \ref{fig:ParaPlotsL}.
\end{enumerate}
These last two points are explained in more detail in the last paragraph of this section.

We consider an initial configuration of two approximate vortices separated by a distance $d=3.0$.
We expect different asymptotic configurations in the cases $\kappa>\kappa_c$ and $\kappa<\kappa_c$. For $\kappa>\kappa_c$ the vortices repel and we expect the flow to move towards a configuration of two 1-vortices with large separation, and therefore we fit using $\E_\infty = 2\E_1$ where $\E_1$ is the energy of a single vortex for the appropriate $\kappa$.
For $\kappa<\kappa_c$ the vortices attract and we expect the flow to move towards a single 2-vortex centered at the midpoint between the two initial vortices, so we fit using $\E_\infty = \E_2$ where $\E_2$ is the energy of a single 2-vortex for each $\kappa$ considered.

The results of the fits to \eqref{FitEqnEarly} and \eqref{FitEqnLate} are summarized in Figure \ref{Fig:TScales}.
In order to separate as cleanly as possible the formation timescale from the interaction timescale, we obtain the former from the region $t \in(0,5)$ and the latter from $t \in(25,60)$. Figure \ref{Fig:TScales}
shows clearly that $T_{\rm short} \ll T_{\rm long}$. We therefore conclude that the vortex formation timescale is generically faster than the vortex interaction timescale. As expected, the interaction timescale diverges at $\kappa_c$ where the vortices do not interact, and decreases as we move away from $\kappa_c$ in either direction, and the vortex interactions become stronger. 
The $\kappa$ dependence of the formation timescale is weaker, which similar to what we found for the 1-vortex case (see Figs. \ref{fig:ParaPlotsE} and \ref{fig:ParaPlotsL}). 
We note also that the numerical values of the formation timescale differ from what was obtained in section \ref{vortex-formation-section}. 
From Fig. \ref{fig:ParaPlotsE} we find $T_{\rm short}\sim 0.18$ for a single vortex centered at the origin, while Fig. \ref{Fig:TScales} shows that for the configuration used in this section $0.3 < T_{\rm short} < 0.8$.
These difference are expected, since different initial conditions are used in the two different calculations.
The important point is that we find consistently $T_{\rm long}\gg T_{\rm short}$.

\begin{figure}
\includegraphics[width=0.8\textwidth]{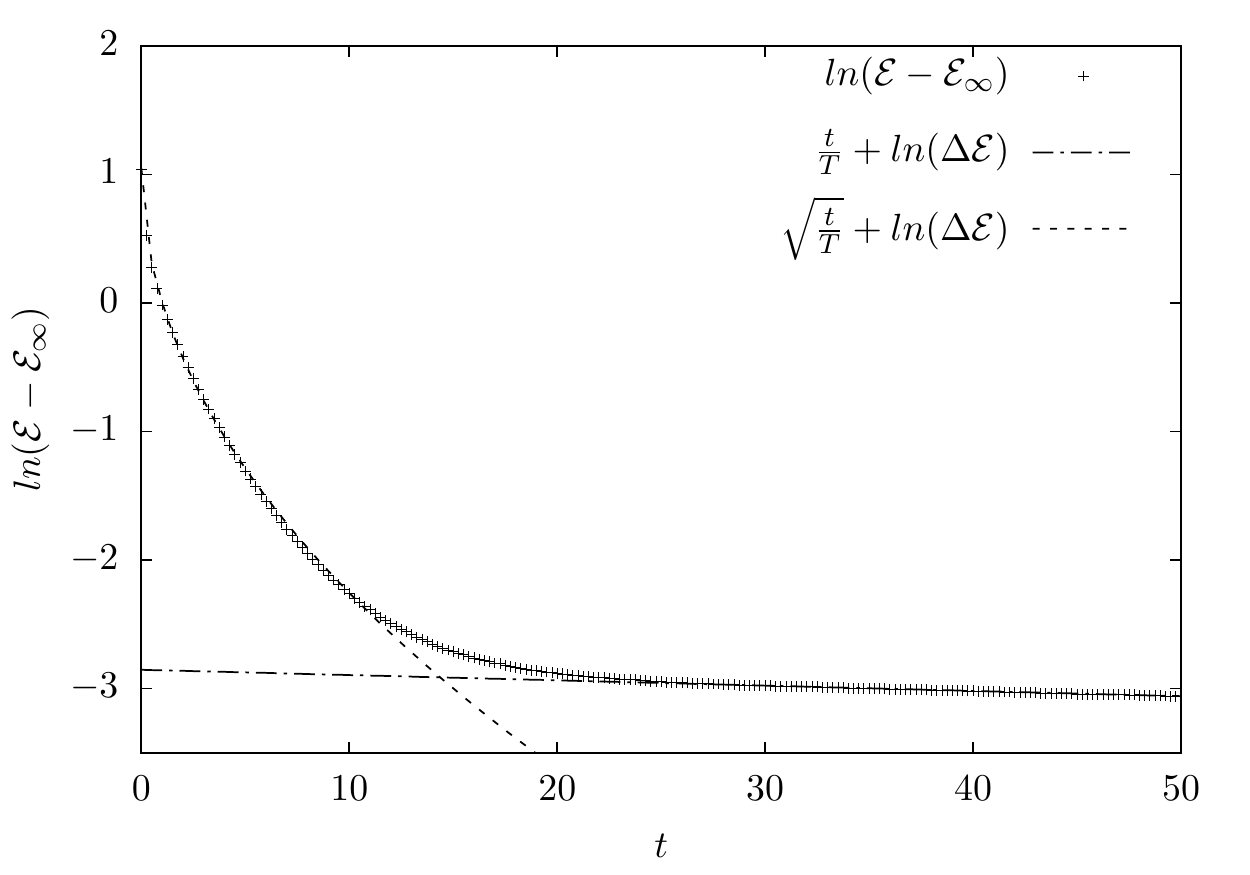}
\caption{\label{fig:2vortexEnergy} $\ln[\E(t)-\E_\infty]$ with $\kappa^2=0.7$ for two approximate 1-vortices which start at $(-1,0)$ and $(1,0)$. The dotted lines are the fits to equations \eqref{FitEqnEarly} and \eqref{FitEqnLate}. (Not all data points are plotted so that the fit lines can be clearly seen.) \label{2scales-fig}}
\end{figure}

\begin{figure}
\includegraphics[width=0.8\textwidth]{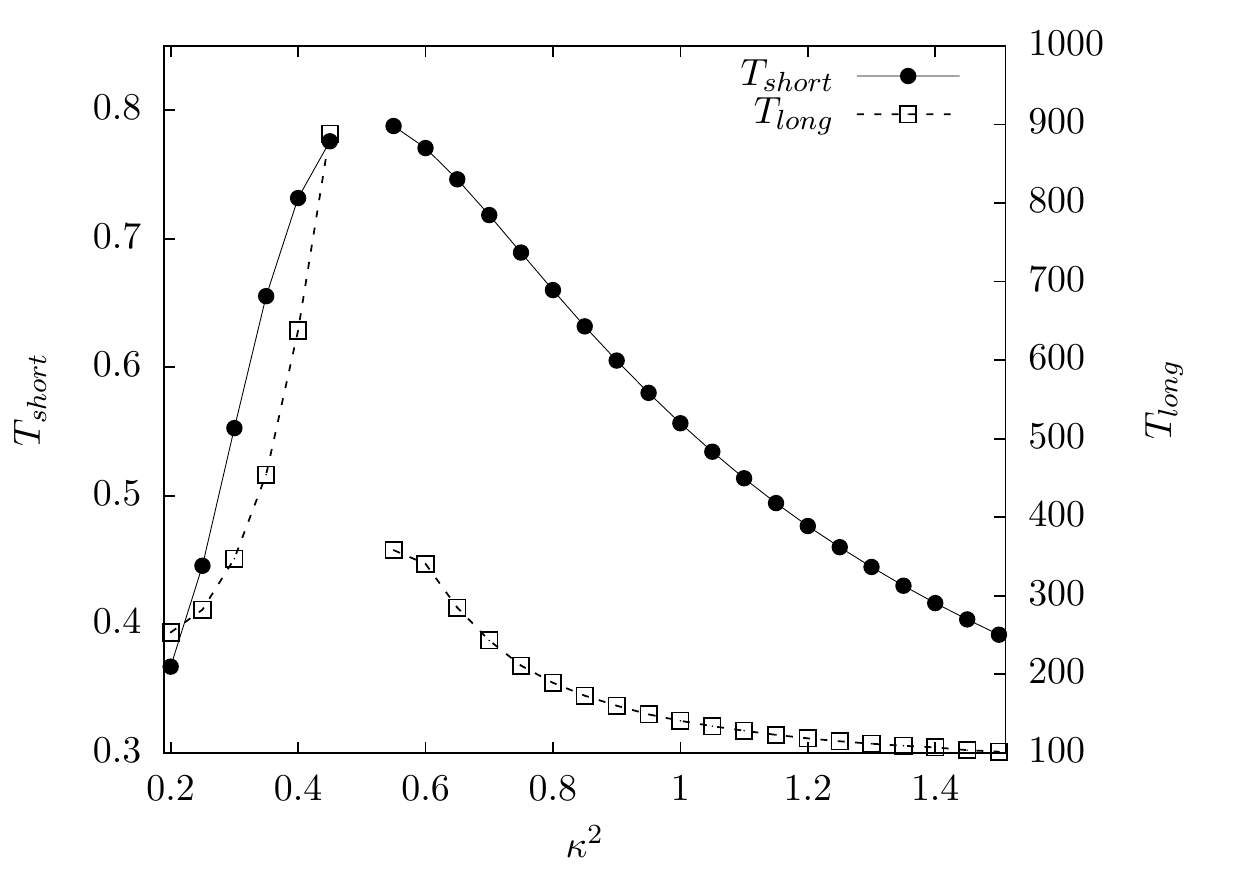}
\caption{\label{Fig:TScales} A plot of the fit parameters $T_{\rm short}$ and $T_{\rm long}$ for several values of $\kappa^2$. Near $\kappa^2_c$ the vortex interactions are very weak and the fits used to determine $T_{\rm long}$ become more uncertain. For $\kappa^2_c$ the vortices do not interact so we exclude this point from the graph.}
\end{figure}

\subsubsection{A 1-vortex  and a (-1)-vortex}
We also look at the interaction between a 1-vortex and a (-1)-vortex. In this case we have an overall winding number of zero, and we expect the two vortices to attract each other and cancel when they merge. In Fig \ref{fig:EnergyOpposite} we plot the energy for $\kappa=\kappa_c$ and $\kappa=1$. We note that vortices with opposite sign will interact in the $\kappa=\kappa_c$ case, and we can clearly see the point where the vortices merge. For $\kappa=1$ the vortices move together rather than apart.
\begin{figure}
\begin{subfigure}[b]{0.49\textwidth}
\includegraphics[width=\textwidth]{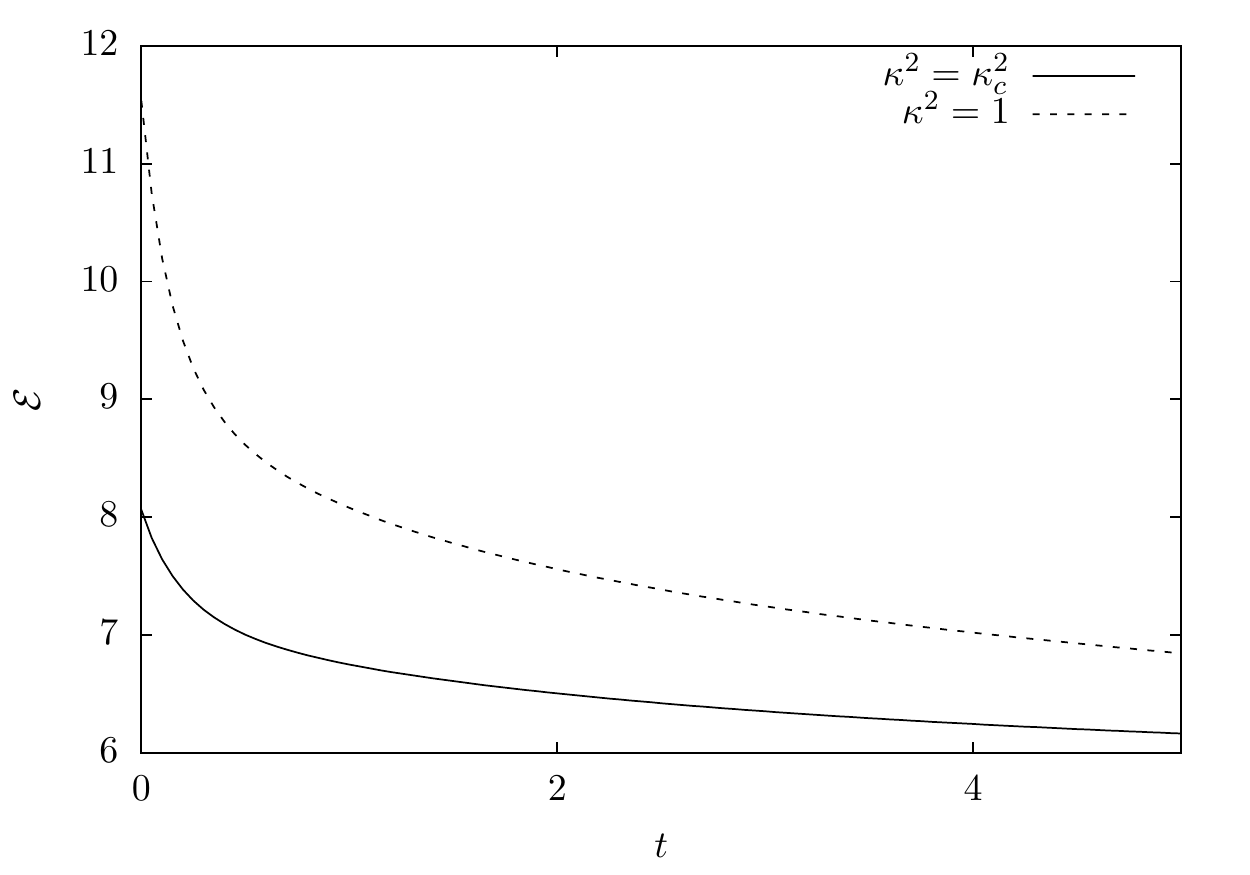}
\caption{The short time behaviour of $\E$.}
\end{subfigure}
\begin{subfigure}[b]{0.49\textwidth}
\includegraphics[width=\textwidth]{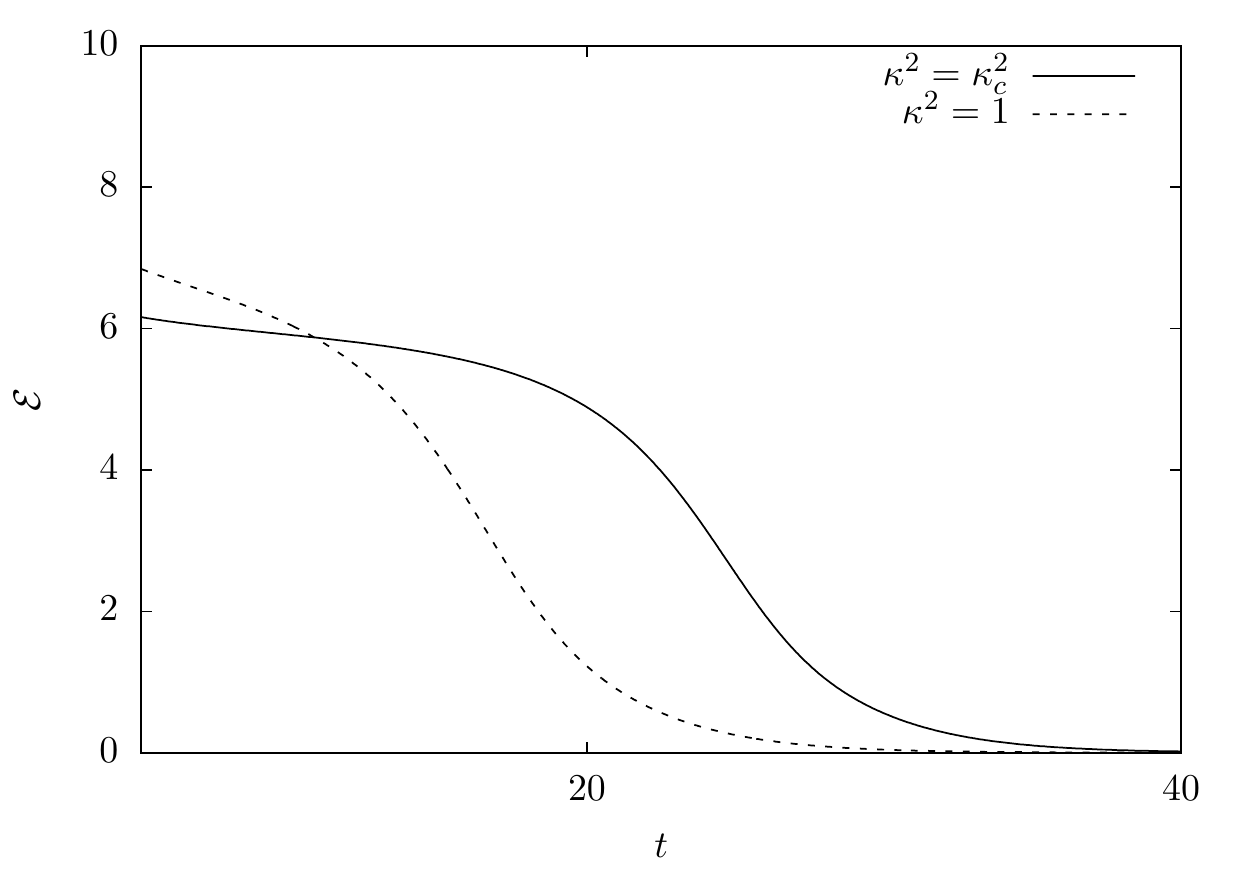}
\caption{The long time behaviour of $\E$. }
\end{subfigure}
\rule{35em}{0.5pt}
\caption{\label{fig:EnergyOpposite} The energy of a 1-vortex and a (-1)-vortex. We start with vortices initially located at $(1.5,0)$ and $(-1.5,0)$, and show $\kappa^2=\kappa^2_c$ and $\kappa^2=1$. For $0<t<5$ we see behaviour similar to the vortex interactions in both cases, but at longer times we see the vortices will attract and merge for any $\kappa$ }
\end{figure}

\subsubsection{Multiple vortices}

Finally, we can also study the energy of multiple vortex configurations as a function of $t$ for different values of $\kappa$. 
We consider two different configurations, both of which have total winding number 4. 
The first configuration is a 3-vortex at (3,0) and a 1-vortex at (-3,0). 
The second has 4 distinct 1-vortices in symmetric locations about the origin: (1,1), (1,-1), (-1,1), (-1,-1). 
The results are shown in Fig. \ref{fig:ConfigEnergy}.
The energy  $\E(t)$ is a monotonically decreasing function.
Initially the energy decreases rapidly as the flow moves from a configuration of approximate initial vortices to true vortices. Then there is a region where the energy evolves  through vortex interactions. The vortices move together or apart depending on the value of $\kappa$. For large $t$ there is an asymptotic region where the vortices have reached a stable configuration either because they are too far apart to interact strongly, or because they have merged into a single vortex. 
From Fig. \ref{fig:ConfigEnergy} it is clear that the two configurations have similar short time behaviour even with different initial energies, but there is a noticeable difference in their behaviour in the region where the 3-vortex is splitting into 3 1-vortices. The asymptotic behaviour is, as expected, the same in both cases.

\begin{figure}
\begin{subfigure}[b]{0.49\textwidth}
\includegraphics[width=\textwidth]{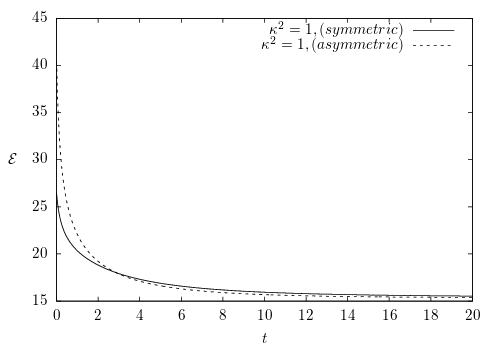}
\caption{The short time behaviour of $\E$.}
\end{subfigure}
\begin{subfigure}[b]{0.49\textwidth}
\includegraphics[width=\textwidth]{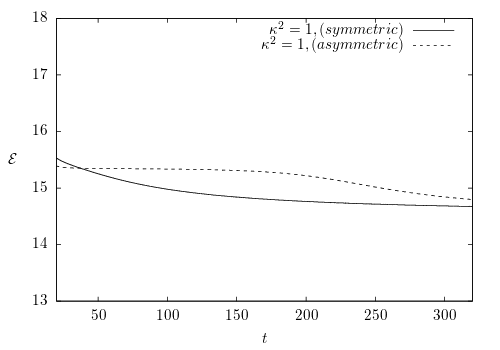}
\caption{The long time behaviour of $\E$. }
\end{subfigure}
\rule{35em}{0.5pt}
\caption{\label{fig:ConfigEnergy} Comparison of the energy of two different configurations with total winding number $n=4$. The symmetric configuration contains 4 1-vortices and the asymmetric one is a 3-vortex in the presence of an additional 1-vortex (see text for details).  The mid range behaviour of $\E$ is altered by the breakup of the 3-vortex in the asymmetric case.}
\end{figure}

\section{Conclusions}
\label{conclusions-section}

We have used the gradient flow equations to study numerically the equilibration of vortices in the Ginzburg-Landau model.  
The flow equations can be used to numerically find  approximate solutions to the GL equations, as the endpoint of the flow, for any number of vortices at arbitrary locations.  
While there are better numerical methods to find solutions, the advantage of our method is that we are able to study the dynamical evolution of the flow.
The primary focus of our work is a study of the timescales associated with vortex formation and vortex-vortex interactions.

We first studied vortex formation by looking at $n=1$ vortices centered at the origin. 
The flow  moves towards vortex solutions as expected, with  winding number corresponding to that of the initial data. We found that this occurred on relatively short timescales that were largely independent of the initial field configuration. 
The vortex formation timescale is proportional to the energy of the vortex, which is a non-linearly increasing function of $\kappa$. 

We also  considered three different kinds of vortex interactions. 
We started with configurations that correspond as closely as possible to two or more separated vortex solutions, for various values of $\kappa$. 
During the early part of the flow the vortex forms rapidly. 
The flow on longer timescales simulates interactions between the vortices as determined by the time dependent Ginzburg-Landau equations. 
We were able to model the attraction between vortices in the   
$\kappa<\kappa_c$ case, and the repulsion in the $\kappa>\kappa_c$ case.
We showed that the interaction timescale decreases when $\kappa$ moves away from $\kappa_c$ in either direction.
We also showed that vortex anti-vortex interactions are attractive for any values of $\kappa$.
Similar results were found in \cite{Chaves} where the authors study the force between vortices by looking at the change in the energy as the separation increases, which is closely related to our gradient flow approach.

Our numerical results raise several interesting questions. 
The first is that the energies in our numerical solutions do not show the $n^2$ scaling for large $\kappa$ that was suggested by approximate solutions in \cite{Nielsen} and the heuristic arguments of \cite{SCText}. 
In addition, it would be good to understand physically the source of the long and short timescales. 
The exponential dependence on ${\sqrt{t}}$ in our fits likely derives from the nonlinear dynamics which affects the early evolution of the far from equilibrium initial configurations. At later times, when the fields are closer to their equilibrium configurations, the dynamics changes qualitatively and linear exponential behaviour is found, as expected.

Several interesting scenarios that have been considered recently using time dependent GL equations of motion could also be studied using a gradient flow approach like the one we have used in this paper. 
Vortices in three dimensional samples constructed from layers which create anisotropy have been studied in a constant external magnetic field in \cite{Liu}, and the dynamics of vortex loops were studied in \cite{Doria,Berd}. Improved Ginzburg Landau models that describe types of multicomponent superconductivity have been studied in \cite{multi1,multi2}. Multiband superconductors \cite{multi3,multi4,multi5} can exhibit new physics and vortex configurations not found in single band superconductors.
 
Finally, we note that while the superconducting behaviour in the Ginzburg-Landau model is due to the symmetry breaking in the scalar field potential, symmetry breaking can occur in more general contexts. Of particular interest is a holographic model of superconductivity \cite{Hartnoll} in which the spacetime around a black hole in Anti-de Sitter space is equivalent to a finite temperature conformal field theory that has been shown to have superconducting properties. Studying the gradient flow (i.e. generalized TDGL equations) in this context could provide insight into the behaviour of these systems away from equilibrium. \\[10pt]
{\bf Acknowledgments}
This research was supported in part by the Natural Sciences and Engineering Research Council of Canada, and a University of Manitoba Graduate Fellowship.

\appendix

\section{Some Numerical Details}
The computations were done using an explicit finite difference scheme for a $20\xi \times20 \xi$ superconductor discretized into a square grid with spacing $\Delta x = \Delta y = 0.05 \xi$.
As one check of convergence we confirmed that our results were not affected by further reducing the grid spacing by half. Specifically, the quality of the fits did not change, nor did the values of the fit parameters to the numerical accuracy of the calculation.

As a further check of the numerical results, we compared the asymptotic value of the $\kappa=\kappa_c$ 1-vortex energy in Fig.~\ref{fig:EPlot} to its known analytic value of $\pi$. We found for large times, namely $t=15$, that $\E(15) = 3.14315391$ and by fitting the long timescale graph we deduced an asymptotic value of energy of $\E(\infty)=3.14145$. These two values differ from the analytic value $\pi$ by  $0.0496\%$ and $0.00454\%$, respectively. 

Data fitting was done using the nonlinear least-squares Marquardt-Levenberg algorithm. In all cases, the $\E(t) \propto e^{-\sqrt{t/T}}$ dependence provided the best fit at the early times, with a reduced chi-squared $\approx 0.01$.

\end{document}